# Integrating systematic surveys with historical data to model the distribution of *Ornithodoros turicata americanus*, a vector of epidemiological concern in North America.

Sebastian Botero-Cañola[1]; Carson Torhorst[1]; Nicholas Canino[1]; Lorenza Beati[2]; Kathleen C. O'Hara[3]; Angela M. James[3]; and Samantha M. Wisely[1]

1-Department of Wildlife Ecology and Conservation, University of Florida, Gainesville, Florida.

2- US National Tick Collection, Institute for Coastal Plain Science, Georgia Southern University, Statesboro, GA 30458, USA

3-USDA, Animal and Plant Health Inspection Service (APHIS), Veterinary Services (VS), Center for Epidemiology and Animal Health (CEAH), Ft. Collins, Colorado 80526

## Abstract

Globally, vector-borne diseases are increasing in distribution and frequency, affecting humans, domestic animals and livestock, and wildlife. Science-based management and prevention of these diseases requires a sound understanding of the distribution and environmental requirements of the vectors and hosts involved in disease transmission. Integrated Species Distribution Models (ISDM) account for diverse data types through hierarchical modeling and represent a significant advancement in species distribution modeling that have not yet been leveraged in disease ecology. We used this approach, as implemented in the recently developed R package RISDM, to assess the distribution of the soft tick subspecies *Ornithodoros turicata americanus*. This tick species is a potential vector of African swine fever virus (ASFV), a pathogen responsible for an ongoing global epizootic that threatens agroindustry worldwide. We created an ISDM for *O. t. americanus,* using systematically collected field data and historical records of this tick species in the southeastern US, to predict its distribution and assess potential correlations with environmental variables. Given the novelty of this method, we compared the results to a conventional Maxent SDM and validated the results through data partitioning using true skills statistics (TSS), sensitivity, and area under the ROC curve (AUC) metrics. Our input for the model consisted of detection data from 591 sampled field sites and 12 historical species records, as well as four variables describing climatic and soil characteristics. We found that a combination of climatic variables describing seasonality and temperature extremes, along with the amount of sand in the soil, determined the predicted intensity of occurrence of this tick species. When projected in geographic space, this distribution model predicted 62% of Florida as suitable habitat for this tick species. The ISDM presented a higher TSS and AUC than the Maxent conventional model, while sensitivity was similar between both models. Our case example shows the utility of ISDMs in disease ecology studies and highlights the broad range of geographic suitability for this important disease vector. These results provide important foundational information to inform future risk assessment work for tick-borne relapsing fever surveillance and potential ASF introduction and maintenance in the US.

**Introduction**

Worldwide, the distribution and frequency of vector-borne diseases are on the rise, affecting humans, domestic animals and livestock, and wildlife (Socha et al., 2022, Swei et al., 2020, Jones et al., 2008). These diseases are expressions of complex systems involving pathogens, hosts, vectors, and the environments that they occupy; an understanding of the ecology of these components and their interactions is of paramount importance in disease management, prevention, and mitigation (Chala and Hamde, 2021). The geographical distribution and habitat requirements of vectors is critical for understanding and predicting the spatiotemporal patterns of disease risk (Johnson et al., 2019). Although the creation of occurrence data repositories (Heinrich et al., 2015; Kraemer et al., 2015), and advances in species distribution modeling have moved forward, along with our understanding and prediction capabilities for many disease systems (Peterson, 2014), there are still many vector-borne disease systems for which the ecology and distribution of the vector are poorly understood. One such group of arthropods are soft ticks (Acari: Argasidae), vectors of pathogenic bacteria and viruses whose nidicolous lifestyle (living in burrows, nests and crevices used by their hosts) has hindered their widespread collection and ecological study (Vial, 2009).

Species distribution models (SDM) are useful tools for understanding and predicting the distribution of the pathogens, hosts, and vectors involved in disease systems, particularly for understudied species such as Argasid ticks (Sage et al., 2017). By relating occurrence data of a species to environmental conditions, these models not only enable researchers to assess the effect of environmental variables on the habitat suitability or abundance of a species, but they also allow the projection of these relationships over current or future geographical space. These capabilities render SDMs powerful tools for public and animal health decision makers (Peterson, 2014). Despite their effectiveness and widespread use, SDM's are subject to biases that need to be properly tested and validated before the information is applied (Soley-Guardia et al., 2024). The most prevalent bias is the non-random collection of occurrence data, ranging from convenience sampling to museum specimens to systematic monitoring, all of which differ in their accuracy and precision in both geographic location and the probability of detection. If not properly accounted for, these issues can significantly impact model results and utility (Kramer-Schadt et al., 2013).

Hence, it can be argued that the next frontier for SDM is the creation of models capable of merging diverse data sources while accounting for the inherent characteristics of each data type. Integrated Species Distribution Models (ISDM) are hierarchical models that allow for the smooth integration of disparate data by modeling first the observation process for each dataset, and then the underlying species presence/abundance process governing all datasets (Fletcher et al., 2019). This recent advance in SDM can greatly improve our ability to use data from disparate sources, account for differences in detection probability, and produce more robust inferences about the distribution of a species. Nonetheless, there are still few case studies that use these methods.

Soft ticks (Argasidae) are a taxonomic group with sparse and disparate occurrence data compared to that of hard ticks (Ixodidae). Argasid ticks of the genus *Ornithodoros* are important components of multiple disease systems, serving as vectors for pathogens that impact both humans and animals globally (Hoogstraal, 1985). For example, these ticks are vectors of various relapsing fever bacteria affecting humans (*Borrelia* sp.; Jakab et al., 2022). In Africa and Europe, *Ornithodoros* ticks are involved in the transmission of African swine fever virus (ASFV), a disease with a complex transmission system that impacts wild and domestic pigs (Galindo & Alonso, 2017; Bonnet et al., 2020) and that has had tremendous impacts on the economy and food security (Chenais et al., 2017; Nguyen-Thi et al., 2019; Stancu, 2019). ASFV can be transmitted directly, environmentally, and via soft tick vectors; the latter transmission pathway can create an endemic sylvatic transmission cycle that is difficult to mitigate and control (Boinas et al., 2011). In North America, *Ornithodoros turicata* is considered a competent vector of this virus (Golnar et al., 2019). Although the virus has not been reported in the United States (US), detection, mitigation, and control plans have been prepared in case of an outbreak (Brown and Bevins, 2018). *Ornithodoros turicata americanus*, the eastern subspecies, known from just a few localities in Florida, is of particular interest given that this state may be at a higher risk for an ASF introduction due to the high densities of feral pigs and its proximity to an ongoing outbreak in the Caribbean (Wormington et al., 2019).

Despite the importance of *O. t. americanus*, little information is available on the distribution of this tick species in the southeastern United States. Donaldson et al. (2016) modeled the distribution of *O. turicata* throughout North America; however, the eastern populations comprised < 8% of the occurrence records, and the authors reported some underprediction by their model in the range of the eastern subspecies, *O turicata americanus*. Furthermore, *O. t. americanus* shows biological and behavioral differences with the western populations of this species (Beck et al., 1986), and hence could have different ecological requirements. A higher-resolution analysis of the distribution of *O. t. americanus* is needed to better understand the transmission risk posed by these ticks in the event of introduction of ASFV to the US, and to inform monitoring and mitigation plans as part of an outbreak response (USDA, 2023).

Considering the need for a higher resolution analysis, and the potential offered by ISDMs, we modeled the distribution and environmental requirements of this tick species using integrated and conventional SDM. In this study, our objectives were to: i) design and conduct a systematic survey of the occurrence of *O. t. americanus* in the southeastern US; ii) create an ISDM using the collected field data and historical occurrence information; iii) given the novelty of this method, evaluate the performance of the model and compare its results to a conventional Maxent SDM; and iv) assess the importance and relationship between environmental variables and *O. t. americanus* distribution. Specifically in this last objective, we want to test two hypotheses related to the specific habitat requirements of Argasid ticks. First, given their nidicolous lifestyle, we hypothesized that soil properties would influence the quality of burrows as microhabitats and hence affect the large-scale distribution of these ticks. Second, we hypothesized that due to the relatively stable and moderate microclimate within burrows and dens, soft ticks would be resistant to daily changes in climatic conditions and mostly influenced by extreme environmental conditions over their life cycle (Sage et al., 2017).

**Methods**

*Study area and environmental variables*

The selection of the modeling area is a key part of the species distribution modeling process and should represent the sites accessible to the species of interest over a defined time period (Barve et al., 2011; Soley-Guardia et al., 2024). We selected the states of Florida, Mississippi, Alabama, Georgia, S. Carolina, and N. Carolina (Figure 1) for the following reasons: i) their proximity to Florida, the only state east of the Mississippi river where *O. turicata* has been detected; ii) shared ecoregions (Omernik and Griffith, 2014); and iii) the presence of suspected or known host species for *O. turicata* (Donaldson et al., 2017).

We selected two sets of environmental variables to model the distribution of *O. t. americanus*. The first set consisted of 20 bioclimatic variables obtained from the BIOS+ dataset of the CHELSA V2.1 initiative (Brun et al., preprint; Brun et al., 2022; Supporting Information S1). These variables describe the inter-annual averages, temporal variation and extremes of temperature, precipitation, humidity, and water balance, which are factors shown to influence the distribution of soft ticks (Estrada-Peña et al., 2010; Donaldson et al., 2017; Sage et al., 2017). Given the nidicolous and burrow-dwelling nature of *O. turicata*, we also hypothesized that soil conditions would influence tick survival and distribution. Therefore, our second dataset, obtained from the SoilGrids database, consisted of seven variables describing soil properties averaged for the top 2 m of soil. This database maps the spatial distribution of soil properties across the globe using machine learning methods and a large dataset of observations (Poggio et al., 2021; Supporting Information S1).

To reduce the high dimensionality of our dataset, we used multivariate methods to extract principal components from the two datasets, climatic or soil conditions. We used the *prcomp* function of the Terra package in R Studio (Hijmans, 2022) to perform a principal component analysis (PCA) of the climatic data set. The first two principal components (PC) cumulatively accounted for 71% of the variation. The values of the first principal component (Clim_PC1) were negatively associated with the seasonality of temperature and precipitation, as well as the magnitude of cold extremes. The second PC (Clim_PC2) was positively associated with water balance and precipitation, and negatively with extreme heat (Figure 2D). Similarly, we created a PCA for the variables describing soil physicochemical characteristics, with the first two PCs accounting for 82% of the variation. In this case, the first PC (Soil_PC1) was inversely related to pH and carbon content, and positively associated with the bulk density of the fine earth fraction. The second PC (Soil_PC2) was positively correlated with the volumetric fraction of coarse material in the soil (Supporting Information S1). Finally, we created a robust PCA of soil composition data (percentages of clay, sand and silt) using the *princomp.acomp* function of the Compositions R package (van den Boogaart et al., 2013), and, as the first PC explained 95% of the variation with a strong negative relation to sand content, we used this PC to describe soil composition (Figure 2E). All raster variables were resampled to a 2x2 km cell size.

We calculated weather and local conditions at the time and place each burrow was sampled (see below). These variables were used to account for sampling artifacts that could influence detection probability of the ticks. The local conditions included the topographic wetness index (TWI) at a resolution of 30 m, and the percentage of tree cover at a resolution of 500 m. We also calculated the temperature, relative humidity, and total precipitation during the five days prior to each sampling event as these variables influenced *O. t. americana* detection probability (Canino et al., in review).

*Field surveys of O. t. americanus*

Due to the lack of georeferenced occurrence data on *O. t. americanus*, we designed and conducted a systematic survey of Florida. To identify the sampling sites, Florida was divided into 10x10 km quadrants, and 102 quadrants were selected using a random stratified strategy. First, we divided the study area by the ecoregions aggregated by Crawford et al. (2020) representing significant biogeographic units, with the number of quadrants to sample from each ecoregion being proportional to its relative area. This approach ensured that the full environmental variation of the study area was sampled. Subsequently, due to access logistics, only quadrants overlapping public lands were considered. Finally, given that *O. t. americanus* has only been found within gopher tortoise burrows, despite sampling efforts in other microhabitats (Adeyeye, 1982), we focused our detection efforts on these burrows. Accordingly, quadrants within each ecoregion where gopher tortoises had previously been detected were randomly selected, with the probability of selection proportional to the number of gopher tortoise burrows present within public lands of each quadrant (burrow location data was provided by Florida Fish and Wildlife Conservation Commission; Fig 1). Eleven more quadrants were included post-hoc to extend the geographic variation covered by our sampling scheme for a total of 113 quadrants.

Using simulations informed by the estimates of detection probability provided by Canino et al. (in review, submitted January 2024), we determined that by surveying five gopher tortoise burrows (burrows henceforth) at each site, we would have an 85% average probability of detecting the species if present (Supporting Information S2). As a result, at each site, we surveyed five burrows, when possible 100 meters apart, using a modified leaf vacuum to extract 1-3 kg of substrate material from the burrow (Canino et al., in review). Vacuuming was performed on the entrance, walls, roof, and base of the burrow down to a depth of one meter within the burrow. The gopher tortoise is categorized as a threatened species in Florida (Enge et al., 2006); the one meter maximum sampling depth was a requirement of our scientific collection permit to minimize the impact of this work on the species. Subsequently, the samples were sieved at the laboratory to extract the soft ticks present in the burrow material (Canino et al., in review). All sampling was conducted on public lands with official permits from the respective managing agency, and under the scientific collection permit of the Florida Fish and Wildlife Conservation Commission (No. LSSC-22-00054).

*Historical records*

We also obtained historical records of *O. t. americanus* from the United States National Tick Collection, The Global Biodiversity Facility (GBIF.org, 2023), Cooley and Kohls (1944), and the original records reported in Donaldson et al. (2017). The records were georeferenced using best practices, and their spatial uncertainty was estimated (Chapman and Wieczorek, 2006). Only records with uncertainty of 15 km or less were included.

To account for biases arising from disparities in sampling effort over the study area (Fourcade et al., 2014), we created a raster describing overall tick sampling intensity. To do this, we downloaded all georeferenced records from Ixodida (hard and soft ticks) from GBIF (GBIF.org, 2023) and estimated the kernel density of these occurrence records over the study area using a sigma=100 km. The values of this raster were then rescaled from 0.1 to 1.

*Integrated Species Distribution Modeling*

We fitted an ISDM using the framework described by Fletcher at al. (2019) and implemented in the R package RISDM (Foster et al., 2023). This hierarchical modeling framework describes the distribution of a species as an inhomogeneous point process, with covariates influencing the point intensity (distribution model). The distribution model is linked to the observed data through different sub-models (observation models) describing how each data type was acquired over the unobserved (latent) point process. In the case of the historical, presence only data (PO), the observed data is modeled as resulting from a sample of the point process that can be biased by a covariate. On the other hand, the systematically collected presence-absence data (PA) are related to the point process by recording the sites where the counts are 0 or greater than 0 using the link between Poisson and Bernoulli random variables (Equation 4 in Fletcher et al., 2019). The PA sub-model can also include sampling artifacts, which are variables that influence the probability of detecting the species. The Foster et al. (2023) model also requires the value of the extent of the sampled area of each unit in the PA dataset to link this sub-model to the occurrence intensity of the species. Finally, this modeling framework allows for the inclusion of a spatial random effect to account for spatial autocorrelation or missing but relevant and spatially-smooth covariates.

As inputs for the model, we used the PA dataset from our systematic field sampling, with each burrow serving as a site at which survey resulted in detection or no detection of the tick. We explored the influence of weather conditions, the topographic wetness index at the burrow, and the percentage of forest cover around it on the probability of detecting the tick. Given that our PA sampling sites consisted of points (burrows), we included the weight of the sand sample extracted from the burrow as an index of survey effort in the area parameter. Including the estimated area searched for finding each burrow yielded almost identical results. For PO data, we included the curated historical records, and the bias layer to parametrize the sampling bias term. Finally, we created a spatial mesh, used for model prediction and spatial effects evaluation, using the *makeMesh* function of the RISDM package setting 4 km as the assumed effective range of the spatial random effect, a maximum number of 2000 nodes, and an expansion amount of 20, with the rest of the parameters kept at their default. The default priors were used in the creation of all models.

With these datasets, we used the function *isdm* (RISDM) to create models exploring first the effect of TWI, forest cover, and climatic conditions on PA detection probability. Then, when the variable best describing sampling artifacts was identified, models for the main climate (Clim_PC1 and Clim_PC2) and soil variables (Soil_PC1, and Soil composition) were explored using linear and quadratic relations. Lastly, a final model was created including the variables found to be significant in the previous step. Two criteria were used to assess each model: only models in which the beta credible intervals did not include 0 were kept, and the marginal likelihood of the model was used to compare models.

The output of point process ISDM is an intensity parameter that is interpreted as indicative of the number of individuals in an area. Because the PA sampling sites in our study consisted of individual burrows, yet the historical data consisted of samples of multiple individuals from a particular burrow, we interpreted the intensity parameter as a relative measure of the number of infested burrows on the landscape, as opposed to the number of individuals, and we called it occurrence intensity.

*Maxent modeling*

We compared the results obtained by the ISDM with a Maxent model, a more conventional approach to SDM. Maxent compares presence records against the environmental background and estimates habitat suitability by finding the distribution of maximum entropy subject to the constraint that the expected value for each environmental variable under this estimated distribution matches its empirical average (Elith et al., 2011). This algorithm has been repeatedly shown to perform well and to display a satisfactory prediction capacity (Elith et al., 2006; Phillips and Dudík 2008). We also selected Maxent due to its ability to explore more complex relationships between the occurrence of the tick species and the covariables, as well as its capacity to include a larger number of covariates. In addition, Maxent permits the inclusion of a bias layer, which is used to extract the background data in proportion to sample intensity, and thus account for sampling disparity over the study area (Phillips et al., 2010).

We combined the historical records and the locations where our field survey detected the ticks as the occurrence input for the Maxent model. For a sampling bias layer, we used the kernel density map describing the distribution of detection records for Ixodida in GBIF GBIF ?(GBIF.org, 2023) and all the sampled field locations included in this study. For environmental covariates we used the first five principal components of climate, the first component of soil composition, and the first two principal components of soil physicochemical characteristics (Figure 2; Supporting information S1).

Maxent can fit very complex models; however, this can result in overfitting (Radosavljevic and Anderson, 2014). We optimized model complexity by selecting a combination of the regularization multiplier (RM) and feature classes (FC) which determine the penalty associated with including variables and the shape of the response curves respectively (Merow et al., 2014). Using the workflow provided by the package ENMeval (Kass et al., 2021), we partitioned the occurrence and background data into four spatial blocks using the *get.checkerboard2* function. Then, using the *ENMevaluate* function, we ran four replicates, leaving one partition out for validation each time, using every combination of predefined RM (1,2,3,5,10,15) and FC (linear, quadratic, product, and hinge – threshold was not included to avoid overly complex and unrealistic models; Phillips et al. 2017). We used a two-step selection process to select an appropriate combination of RM and FC. First, we selected the models presenting the lowest (> 10[th] percentile) omission on training data (or.10p.avg), and the highest average Boyce index (>90[th]), and then, selected the model with the lowest AICc among the subset candidates. Finally, we ran 50 bootstrap replicates of the model using the tuned parameters and leaving a random 20% of the occurrence records to map model uncertainty.

*Model evaluation and variable response*

We estimated the occurrence intensity of *O. t. americanus* over the study area by calculating the median of the predicted intensity value from 800 draws of the parameter posterior distributions of the final ISDM model. To estimate a distribution area from this continuous intensity value, we selected a threshold that resulted in a 2% omission rate over the systematically collected presence data. We also used the predictions of 800 draws of the posterior distribution to create the response curves of intensity in relation to the selected variables.

We used the median suitability obtained from the 50 Maxent replicates and estimated the distribution area from this continuous suitability value. We selected a threshold that resulted in a 2% omission rate over the pooled occurrence dataset. We also reported the variable response estimated by the model as the marginal response curves provided by the Maxent software output.

Finally, we evaluated and compared the performance of the ISDM and Maxent models using three metrics: i) the True Skills Statistics (TSS), assessing the ability of the reclassified model to correctly predict presence and absence of species; ii) sensitivity, evaluating the capacity of the reclassified model to correctly predict the occurrence of the species; and iii) the area under the receiver-operator curve (AUC), a threshold-independent metric widely used for assessing the discriminatory capacity of SDMs that is most informative with presence-absence data and when predicting the realized distribution is the main objective (Mouton et al., 2010; Jiménez-Valverde, 2012)

To estimate these metrics for each model we first split the PA dataset by partitioning the occurrence and background data into four spatial blocks to create a double checkboard pattern of 20x20 km quadrants immersed in 100x100 km cells using the *get.checkerboard2* function (Kass et al., 2021; Supporting Information S3). This data partition strategy was used to minimize the biasing effect of spatial correlation between training and testing occurrence data (Muscarella et al., 2014). Then, for each model we created ten replicates using combinations of two and three partitions and leaving the remaining data for model evaluation. As a way of accounting for imperfect detection for model validation, we aggregated the results from the five burrows surveyed at each locality, using the centroid of the burrows as the validation point, and defining it as an occurrence if at least one burrow was positive for ticks, and as an absence otherwise. Each replicate was reclassified using the same approach as for the complete model, and the TSS and sensitivity were calculated from the confusion matrix obtained with the validation dataset. The AUC was estimated using the validation dataset employing the *evaluate* function of the package dismo (Hijmans et al., 2017). A paired t-test was used to assess difference in evaluation metrics between the ISDM and Maxent models.

**Results**

*O. t. americanus data records*

We sampled a total of 591 burrows at 113 sampling sites, of which 203 burrows (34%) from 61 sites (54%) were positive for the presence of *O. t. americanus*. Our review of historical records yielded 19 records for the study area, of which 12 had the required spatial resolution (Figure 1).

*Integrated species distribution model*

The linear coefficients for the effect of relative humidity, precipitation, and forest cover on the detection probability for the PA sub-model were significant. Nonetheless, the model presenting constant detection probability displayed the highest marginal likelihood and was used for the remaining model fitting process (Table 1). In relation to the distribution component of the ISDM, the quadratic effects of Clim_PC1 and Clim_PC2, as well as soil composition, were found to be significant (Table 2). These models supported a significant effect of sampling bias over the observed pattern. When the three quadratic effects were included in the final model, one coefficient for Clim_PC1 overlapped 0 (Table 2). For this reason, and after exploring different combinations of linear and quadratic relationships, we selected the model including a quadratic effect for Clim PC1 and Clim PC2, and a linear relationship for soil composition. Under this model, the effect of sampling bias became insignificant. This model presented the higher marginal likelihood for all explored models (Table 2).

Under the selected model, occurrence intensity in relation to Clim_PC1 peaked at positive values, while decreasing at the extremes of this variable, indicating that *O. t. americanus* reached its highest intensity of occurrence at mildly seasonal climates with modest cold extremes and decreased occurrence intensity at variables representing a tropical climate (Figure 2A). On the other hand, occurrence intensity peaked at the lower end of Clim_PC2, suggesting a positive effect of temperature extremes with a negative influence of precipitation (Figure 2B). The negative effect of soil composition indicated higher suitability when there was a larger proportion of sand in the soil (Figure 2C).

Under this model, the predicted occurrence included Florida and southern Georgia, with north-central Florida presenting a block of high predicted occurrence intensity (Figure 3A). When reclassified, a total area of 109,316 km$^2$ was predicted to be suitable, accounting for 11% of the modeling area. In Florida, 96,316 km$^2$ were predicted suitable for this tick species, representing 62% of the terrestrial area of the state. The southeastern portion of Georgia was predicted to host 12,756 km$^2$ of suitable habitat, representing 8% of the area of the state (Figure 3B).

*Maxent model*

In agreement with the ISDM, the most important variables for training the Maxent model were soil composition and Clim_PC1, with percent contributions of 51% and 43% respectively (Supporting Information S4). The marginal response curves of these two variables displayed a similar shape to those of the ISDM. On the other hand, in the Maxent model Clim_PC2 displayed a 0.4% contribution. The marginal response curve of this variable, peaking and stabilizing at positive values, differed from the results of the ISDM (Supporting Information S4).

*Model comparison and evaluation*

The spatial pattern of suitability predicted by the Maxent model was analogous to ISDM (Figure 3C-D). The main spatial differences among the models consisted of an extended range of suitable habitat in southwestern Florida predicted by the Maxent model, while the ISDM model extended the range of suitable habitat along the southern coast of Georgia (Figure 4).

Regarding the evaluation metrics computed from the confusion matrix (i.e. using the threshold map), the TSS statistic was significantly higher for the ISDM at an average of 0.36 compared to 0.29 for the Maxent model (p-value=0.05). There were no significant differences in the sensitivity between both models (0.92 and 0.93 for the ISDM and Maxent model respectively; p-value=0.83). The AUC value of the models showed a marked difference, with an average value of 0.81 for the ISDM and 0.65 for the Maxent model (p-value=0.0002; Figure 5).

**Discussion**

Our study demonstrates the value of using a robust statistical framework integrating disparate occurrence data to predict the distribution of a species of medical and veterinary health concern. Using ISDM allowed for a greater understanding of the biases associated with our sampling methods, while also providing insights into the responses of *O.t. americanus* to the environmental variation outside the area where the systematic sampling was conducted. Furthermore, model evaluation suggests that ISDM has a higher predictive performance than a Maxent model. Regardless of the modeling method used to describe the argasid's distribution, our results highlight that *O. t. americanus* is widely distributed throughout Florida. Given the risk factors for ASFV introduction into the state (USDA 2018), the potential for *O. t. americanus* to contribute to transmission and endemicity of ASFV in the region requires further evaluation.

*Insights into the ecology of O. t. americanus*

The ISDM approach provided a method to explicitly test hypotheses regarding the influence of environmental covariates on the occurrence of *O. t. americanus*. Soil composition was a key variable influencing the occurrence of this soft tick in the southeastern US. This work supports the potential of the SoilGrids dataset to model burrowing or subterranean species (Austrich, 2021; Gardner et al., 2021; Pavleck et al., 2021). We speculate that the strong effect of soil composition on *O. t. americanus* occurrence could be because sandy soils are better drained, protecting tick microhabitat from flooding or dampening conditions which have been shown to be detrimental for the ticks (Adeyeye and Butler, 1989). This finding supports the additional hypothesis that climatic extremes have more influence than averages (Sage et al., 2017).

It is noteworthy that the distribution models, as well as the PA data, differentiated the distribution of the *O. t. americanus* from the distribution of its habitat host, the gopher tortoise (Figure 1).

This disparity in distribution suggests that the species has a unique set of environmental requirements, contained within but different from those of the gopher tortoise. Studies assessing the environmental niche relations of species involved in close ecological relationships such as parasitic plants and their hosts, or plague and its mammalian hosts, have obtained similar results in which the parasite species displays a unique niche contained within that of its hosts (Maher et al., 2010; Lira-Noriega et al., 2014). Nonetheless, our distribution models do not predict the occurrence of suitable habitat outside the distribution of the gopher tortoise, indicating that another habitat host may not play an important role in extending the distribution of the species. However, further studies are needed to determine if any other species may contribute to the maintenance of the tick species within its distributional area.

A large proportion of the area of Florida provides suitable habitat for *O. t. americanus* with more than 50% of the visited locations presenting infested burrows, yielding an average of 15 and maximum of 232 recovered ticks. The ubiquity of this competent vector, coupled with high densities and frequent legal and illegal movement of feral swine in Florida (Lewis et al., 2017; Hernández et al., 2018), highlights the need for epidemiological modeling of the potential role of *O. t. americanus* in the sylvatic maintenance of ASFV in Florida.

*ISDM considerations*

The use of an ISDM approach greatly increased our capacity to explore a wider environmental gradient beyond that of our systematically sampled sites (Fletcher et al., 2016). Contrary to other examples of PO and PA data integration (Fletcher et al., 2016; Mäkinen et al., 2022), in our study the systematically collected data contained more records and covered a larger geographical range. Thus, an ISDM for these datasets increased the background area against which the intensity of the point process model is estimated and the effect of the covariables is assessed. This expansion resulted in a model that, while leveraging the more informative PA dataset, also incorporated a wide environmental gradient and ultimately displayed higher predictive capacity than the PO Maxent model. The fact that both the ISDM and Maxent models displayed the same capacity to correctly predict the occurrence of this tick species (sensitivity), while the ISDM outperformed the Maxent model in metrics incorporating specificity, suggests that the power of the ISDM in this study is the reduction in overpredicting areas where the ticks are absent.

Our field survey scheme focused on sampling available burrows dispersed over a variable area of land at each location, rather than surveying a defined area, hindering the clear translation of the intensity parameter to an abundance or density estimate. Hence, our intensity function can be interpreted as a relative index of the occurrence of infested burrows over the landscape. With the modeling approach and interpretation taken here, variation in the abundance of ticks within a burrow is not included in the modeling process. Future studies may consider exploring this aspect, as it could have an important impact on disease transmission risk.

Although the PA sub-models indicated that the climatic conditions and landscape structure influenced the detectability of the soft ticks within the burrows, the constant detectability sub-model was found to be the best fitting. This indicated that overall, these effects were small, attesting to the efficacy of the burrow vacuum method to collect soft ticks under different environmental conditions. Our results provided a justification for our use of PA to estimate the ROC curve, as non-detections could be interpreted as unbiased indicators of absences (MacKenzie et al., 2002). Similarly, the PO sub-model indicated support for an effect of collection bias on the observed intensity distribution with simpler models, yet the final model including the three covariates did not support a bias effect. This finding suggests that the data were not biased in relation to the set of environmental variables, and/or that spatial sampling bias of soft ticks might not be captured by the sampling intensity layer constructed from records predominantly of hard ticks.

*ISDM as tools for disease geography*

The rapid pace of global change means that researchers and public health officials require fast and reliable methods to understand the distribution and responses of vectors, hosts, and pathogens involved in disease systems. ISDMs appear to be an effective way to leverage disparate data to gain insights into these complex disease systems. Our study demonstrates that explicitly including sub-models for different types of data provided more robust prediction of the distribution of *O. t. americanus* than with an approach consisting of just pooling disparate data together. Furthermore, these types of models allow for the integration of abundance data to estimate the relative abundance of a species over space (Morera-Pujol et al., 2023), a variable that can be extremely important for epidemiological modeling and forecasting (Acevedo et al., 2007; Barasona et al., 2014; Tkadlec et al., 2019).

## Data availability

Given that the Gopher tortoise is listed as a Threatened species in Florida, we agreed with the Florida Wildlife Commission not to make publicly available the locations of sampled burrows. The rasters containing the predicted occurrence intensity from the ISDM, as well as the predicted suitability from the Maxent model are available through Figshare (NNNNN).

## Acknowledgements

This work was funded by cooperative agreements (CA#: AP23VSSP0000C117, CA#: AP22VSSP0000C050) with the United States Department of Agriculture, Center for Epidemiology and Animal Health and by the USDA National Institute of Food and Agriculture, McIntire-Stennis project 7004318. The findings and conclusions in this document are those of the authors and should not be construed to represent any official USDA or U.S. Government determination or policy. We are grateful to Dr. Scott D. Foster from Data61, CSIRO, for his technical support with the RISDM package; and Dr. Benjamin Baiser for his advice on environmental data processing. In addition, we thank Zoe White, Rayann Dorleans, Elizabeth Garcia, Kristen Wilson, Madison Heisey, and Lucy Baptist for their help in the field, laboratory, and data processing components of this project. We are also grateful to Nicole DeSha from the Florida Fish and Wildlife Conservation Commission for the support at various stages of this project. Finally, we would like to thank the managing agencies and staff that kindly provided us with access and logistic support to conduct research at our sampling locations.

## References

Acevedo, P., Vicente, J., Höfle, U., Cassinello, J., Ruiz-Fons, F. and Gortázar, C. 2007. Estimation of European wild boar relative abundance and aggregation: a novel method in epidemiological risk assessment. Epidemiology & Infection, 135: 519-527.

Adeyeye, O. A. 1982. Field studies on *Ornithodoros turicata* duges in the gopher tortoise (*Gopherus polyphemus* Daudin) habitat in north central Florida. M.S. Thesis, University of Florida, Gainesville, Florida, USA. 109pp.

Adeyeye, O.A., and Butler, J.F. 1989. Population structure and seasonal intra-burrow movement of *Ornithodoros turicata* (Acari: Argasidae) in gopher tortoise burrows. Journal of medical entomology, 26: 279-283.

Austrich, A., Kittlein, M. J., Mora, M. S., and Mapelli, F. J. 2021. Potential distribution models from two highly endemic species of subterranean rodents of Argentina: which environmental variables have better performance in highly specialized species?. Mammalian Biology, 101:503-519.

Barasona, J.A., Mulero-Pázmány, M., Acevedo, P., Negro, J.J., Torres, M.J., Gortázar, C. and Vicente, J., 2014. Unmanned aircraft systems for studying spatial abundance of ungulates: relevance to spatial epidemiology. PloS one, 9:115608

Barve, N., Barve, V., Jiménez-Valverde, A., Lira-Noriega, A., Maher, S. P., Peterson, A. T., Soberón, J., and Villalobos, F. 2011. The crucial role of the accessible area in ecological niche modeling and species distribution modeling. Ecological Modeling 222:1810-1819.

Beck, A., Kenneth F., Holscher, H., and Butler, J.F. 1986. Life cycle of *Ornithodoros turicata americanus* (Acari: Argasidae) in the laboratory. Journal of medical entomology, 23: 313-319.

Boinas, F.S., Wilson, A.J., Hutchings, G.H., Martins, C., and Dixon, L.J. 2011. The persistence of African swine fever virus in field-infected *Ornithodoros erraticus* during the ASF endemic period in Portugal. PloS one, 6: p.e20383.

Bonnet, S.I., Bouhsira, E., De Regge, N., Fite, J., Etoré, F., Garigliany, M.M., Jori, F., Lempereur, L., Le Potier, M.F., Quillery, E. and Saegerman, C. 2020. Putative role of arthropod vectors in African swine fever virus transmission in relation to their bio-ecological properties. Viruses, 12: p.778.

Brown, V. R., and Bevins, S.N. 2018. A review of African swine fever and the potential for introduction into the United States and the possibility of subsequent establishment in feral swine and native ticks. Frontiers in veterinary science 5(2018): 11.

Brun, P., Zimmermann, N.E., Hari, C., Pellissier, L., Karger, D.N. (preprint): Global climate-related predictors at kilometre resolution for the past and future. *Earth Syst. Sci. Data Discuss*. https://doi.org/10.5194/essd-2022-212

Brun, P., Zimmermann, N.E., Hari, C., Pellissier, L., Karger, D. 2022: Data from: CHELSA-BIOCLIM+ A novel set of global climate-related predictors at kilometre-resolution. **EnviDat.** https://doi.org/10.16904/envidat.332

Chala, B. and Hamde, F., 2021. Emerging and re-emerging vector-borne infectious diseases and the challenges for control: a review. Frontiers in public health, 9: p.715759.

Chapman, A. D., and Wieczorek, J. (eds.). 2006. Guide to Best Practices for Georeferencing. Copenhagen: Global Biodiversity Information Facility.

Chenais, E., Boqvist, S., Emanuelson, U., von Brömssen, C., Ouma, E., Aliro, T., Masembe, C., Ståhl, K. and Sternberg-Lewerin, S., 2017. Quantitative assessment of social and economic impact of African swine fever outbreaks in northern Uganda. Preventive Veterinary Medicine, 144:134-148.

Cooley, R. A, and Kohls, G.L. 1944. The Argasidae of North America, Central America and Cuba. The American Midland Naturalist Monograph No. 1.

Crawford, B. A., Maerz, J. C., and Moore, C. T. 2020. Expert-informed habitat suitability analysis for at-risk species assessment and conservation planning. Journal of Fish and Wildlife Management, 11: 130-150.

Elith, J., Graham, C.H., Anderson, R. P., Dudík, M., Ferrier, S., Guisan, A., Hijmans, R.J., Huettmann, F., Leathwick, J.R., Lehmann, A., Li, J., Lohmann, L.G., Loiselle, B.A., Manion, G., Moritz, C., Nakamura, M., Nakazawa, Y., Overton, J.M.M., Peterson, A.T., Phillips, S. J., Richardson, K., Scachetti-Pereira, R., Schapire, R.E., Soberón, J., Williams, R.E., Wisz, M.S. and Zimmermann, N.E. 2006. Novel methods improve prediction of species' distributions from occurrence data. Ecography 29:129-151.

Elith, J., Phillips, S.J., Hastie, T., Dudík, M., Chee, Y.E., and Yates, C.J. 2011. A statistical explanation of MaxEnt for ecologists. Diversity and Distributions, 17:43-57.

Enge, K. M., J. E. Berish, R. Bolt, A. Dziergowski, and H. R. Mushinsky. 2006. Biological status report - gopher tortoise. Florida Fish and Wildlife Conservation Commission, Tallahassee, USA. 60pp.

Estrada-Peña, A., Nava, S., Horak, I. G., and Guglielmone, A. A. 2010. Using ground-derived data to assess the environmental niche of the spinose ear tick, *Otobius megnini*. Entomologia experimentalis et20his20catea, 137: 132-142.

Fletcher, R.J., McCleery, R.A., Greene, D.U. and Tye, C.A. 2016. Integrated models that unite local and regional data reveal larger-scale environmental relationships and improve predictions of species distributions. Landscape Ecology, 31:1369-1382.

Fletcher, R.J., Hefley, T.J., Robertson, E.P., Zuckerberg, B., McCleery, R.A., and Dorazio, R.M. 2019. A practical guide for combining data to model species distributions. Ecology, 100: e02710.

Foster, S.D., Peel, D., Hosack, G.R., Hoskins, A., Mitchell, D.J., Proft, K., Yang, W., Uribe-Rivera, D.E., and Froese, J.E. 2023. 'RISDM ': species distribution modeling from multiple data sources in R. Ecography (2023): e06964.

Fourcade, Y., Engler, J.O., Rödder, D., and Secondi, J. 2014. Mapping species distributions with MAXENT using a geographically biased sample of presence data: A performance assessment of methods for correcting sampling bias. pLoS ONE 9:1-13.

Gardner, S. L., Botero-Cañola, S., Aliaga-Rossel, E., Dursahinhan, A. T., and Salazar-Bravo, J. 2021. Conservation status and natural history of *Ctenomys*, tuco-tucos in Bolivia. Therya, 12: 15-36.

GBIF.org. 2023. 26 September 2023 GBIF Occurrence Download
https://doi.org/10.15468/dl.wr7yuy

Heinrich, P. L., Gilbert, E., Cobb, N. S., and Franz, N. 2015. Symbiota collections of arthropods network (SCAN): A data portal built to visualize, manipulate, and export species occurrences. [Dataset].

Hernández, F.A., Parker, B.M., Pylant, C.L., Smyser, T.J., Piaggio, A.J., Lance, S.L., Milleson, M.P., Austin, J.D. and Wisely, S.M. 2018. Invasion ecology of wild pigs (Sus scrofa) in Florida, USA: the role of humans in the expansion and colonization of an invasive wild ungulate. Biological Invasions 20: 1865-1880.

Hijmans, R. J., Phillips, S., Leathwick, J., and Elith, J. 2017. Package 'dismo'.–R package ver. 1.1-4.

Hijmans, R. J. 2022. *Terra: Spatial Data Analysis*. https://rspatial.org/terra/.

Hoogstraal, H. 1985. Argasid and nuttalliellid ticks as parasites and vectors. Advances in parasitology, 24: 135-238.

Jakab, Á., Kahlig, P., Kuenzli, E., and Neumayr, A. 2022. Tick borne relapsing fever-a systematic review and analysis of the literature. pLoS neglected tropical diseases, 16: e0010212.

Jiménez-Valverde, A. 2012. Insights into the area under the receiver operating characteristic curve (AUC) as a discrimination measure in species distribution modeling. Global Ecology and Biogeography, 21: 498-507.

Johnson, E. E., Escobar, L. E., and Zambrana-Torrelio, C. 2019. An ecological framework for modeling the geography of disease transmission. Trends in ecology & evolution, 34: 655-668.

Kass, J. M., Muscarella, R., Galante, P.J., Bohl, C.L., Pinilla-Buitrago, G.E., Boria, R.A., Soley-Guardia, M., and Anderson, R.P. 2021. ENMeval 2.0: Redesigned for customizable and reproducible modeling of species' niches and distributions. Methods in Ecology and Evolution, 12: 1602-1608.

Kramer-Schadt, S., Niedballa, J., Pilgrim, J.D., Schröder, B., Lindenborn, J., Reinfelder, V., Stillfried, M., Heckmann, I., Scharf, A.K., Augeri, D.M. and Cheyne, S.M., 2013. The importance of correcting for sampling bias in MaxEnt species distribution models. Diversity and distributions, 19:1366-1379.

Kraemer, M. U., Sinka, M. E., Duda, K. A., Mylne, A., Shearer, F. M., Brady, O. J., ... and Hay, S. I. 2015. The global compendium of *Aedes aegypti* and *Ae. albopictus* occurrence. Scientific data, 2: 1-8.

Lewis, J.S., Farnsworth, M.L., Burdett, C.L., Theobald, D.M., Gray, M., and Miller, R.S. 2017. Biotic and abiotic factors predicting the global distribution and population density of an invasive large mammal. Scientific reports, 7: 44152.

Lira-Noriega, A., and Peterson, A.T. 2014. Range-wide ecological niche comparisons of parasite, hosts and dispersers in a vector-borne plant parasite system. Journal of biogeography, 41: 1664-1673.

MacKenzie, D.I., Nichols, J.D., Lachman, G.D., Droege, S., Royle, J.A., and Langtimm, C.A. 2002. Estimating site occupancy rates when detection probabilities are less than one. Ecology, 83: 2248-2255.

Maher, S. P., Ellis, C., Gage, K. L., Enscore, R. E., and Peterson, A. T. 2010. Range-wide determinants of plague distribution in North America. The American journal of tropical medicine and hygiene, 83: 736.

Mäkinen, J., Merow, C., and Jetz, W. 2024. Integrated species distribution models to account for sampling biases and improve range-wide occurrence predictions. Global Ecology and Biogeography, 33: 356-370.

Merow, C., Smith, M.J., Edwards, T.C., Guisan, A., McMahon, S.M., Wilfried, S.N., Wüest, T.R.O., Zimmermann, N.E., and Elith, J. 2014. What do we gain from simplicity versus complexity in species distribution models?. Ecography 37: 1267-1281.

Morera-Pujol, V., Mostert, P.S., Murphy, K.J., Burkitt, T., Coad, B., McMahon, B.J., Nieuwenhuis, M., Morelle, K., Ward, A.I., and Ciuti, S. 2023. Bayesian species distribution models integrate presence-only and presence–absence data to predict deer distribution and relative abundance. Ecography, no. 2 (2023): e06451.

Mouton, A. M., De Baets, B., and Goethals, P. L. 2010. Ecological relevance of performance criteria for species distribution models. Ecological modelling, 221: 1995-2002.

Muscarella, R., Galante, P. J., Soley-Guardia, M., Boria, R. A., Kass, J. M., Uriarte, M., and Anderson, R. P. 2014. ENM eval: An R package for conducting spatially independent evaluations and estimating optimal model complexity for Maxent ecological niche models. Methods in ecology and evolution, 5: 1198-1205.

Nguyen-Thi, T., Pham-Thi-Ngoc, L., Nguyen-Ngoc, Q., Dang-Xuan, S., Lee, H.S., Nguyen-Viet, H., Padungtod, P., Nguyen-Thu, T., Nguyen-Thi, T., Tran-Cong, T. and Rich, K.M. 2021. An assessment of the economic impacts of the 2019 African swine fever outbreaks in Vietnam. Frontiers in veterinary science, 8: p.686038.

Omernik, J.M. and Griffith, G.E.. 2014. Ecoregions of the conterminous United States: evolution of a hierarchical spatial framework. Environmental Management 54:1249-1266.

Pavlek, M., and Mammola, S. 2021. Niche-based processes explaining the distributions of closely related subterranean spiders. Journal of Biogeography, 48: 118-133.

Poggio, L., de Sousa, L. M., Batjes, N. H., Heuvelink, G. B. M., Kempen, B., Ribeiro, E., and Rossiter, D. 2021. SoilGrids 2.0: producing soil information for the globe with quantified spatial uncertainty, SOIL, 7, 217–240, 2021.

Peterson, A. T. 2014. Mapping disease transmission risk: enriching models using biogeography and ecology. JHU Press.

Phillips, S. J., and Dudík, M. 2008. Modeling of species distributions with Maxent: new extensions and a comprehensive evaluation. Ecography, 31:161-175

Phillips, S.J., Dudík, M., Elith, J., Graham, C.H., Lehmann, A., Leathwick, J., and Ferrier, S. 2009. Sample selection bias and presence-only distribution models: implications for background and pseudo-absence data. Ecological applications, 19: 181-197.

Phillips, S.J., Anderson, R.P., Dudík, M., Schapire, R.E. and Blair, M.E., 2017. Opening the black box: An open-source release of Maxent. Ecography, 40:887-893.

Radosavljevic, A., and Anderson, R.P. 2014. Making better Maxent models of species distributions: complexity, overfitting and evaluation. Journal of biogeography, 41: 629-643.

Sage, K. M., Johnson, T. L., Teglas, M. B., Nieto, N. C., and Schwan, T. G. 2017. Ecological niche modeling and distribution of Ornithodoros hermsi associated with tick-borne relapsing fever in western North America. pLoS neglected tropical diseases, 11: e0006047.

Socha, W., Kwasnik, M., Larska, M., Rola, J., and Rozek, W. 2022. Vector-borne viral diseases as a current threat for human and animal health—One Health perspective. Journal of Clinical Medicine, 11: 3026.

Soley-Guardia, M., Alvarado-Serrano, D. F., and Anderson, R. P. 2024. Top ten hazards to avoid when modeling species distributions: a didactic guide of assumptions, problems, and recommendations. Ecography: e06852.

Stancu, A., 2019. ASF evolution and its economic impact in Europe over the past decade. The USV Annals of Economics and Public Administration, 18(2 (28)), pp.18-27.

Swei, A., Couper, L. I., Coffey, L. L., Kapan, D., and Bennett, S. 2020. Patterns, drivers, and challenges of vector-borne disease emergence. Vector-Borne and Zoonotic Diseases, 20:159-170.

Tkadlec, E., Václavík, T. and Široký, P. 2019. Rodent host abundance and climate variability as predictors of tickborne disease risk 1 year in advance. Emerging Infectious Diseases, 25:p.1738.

USDA. 2018. National Feral Swine Damage Management Program Five Year Report 2014-2018. United States Department of Agriculture, Animal and Plant Health Inspection Service. <https://www.aphis.usda.gov/wildlife_damage/feral_swine/pdfs/nfsp-five-year-report.pdf>.

USDA. 2023. ASF Response Plan: The Red Book <https://www.aphis.usda.gov/sites/default/files/asf-responseplan.pdf>.

van den Boogaart, K. G., Tolosana, R., Bren, M., and van den Boogaart, M. K. G. 2013. Package 'compositions'. Compositional data analysis Ver, 1, 40-1.

Vial, L. 2009. Biological and ecological characteristics of soft ticks (Ixodida: Argasidae) and their impact for predicting tick and associated disease distribution. Parasite, 16: 191-202.

Wormington, J.D., Golnar, A., Poh, K.C., Kading, R.C., Martin, E., Hamer, S.A. and Hamer, G.L. 2019. Risk of African swine fever virus sylvatic establishment and spillover to domestic swine in the United States. Vector-Borne and Zoonotic Diseases, 19:506-511.

# Tables

**Table 1.** Models exploring the effect of sampling artifacts on the systematically collected presence-absence dataset. Poly refers to the quadratic function for the respective variable. The log of the marginal likelihood of each model is reported (Marg.Like), as well as the estimated credible intervals for each coefficient for the covariable (coeff1, coeff2). Coefficients that do not overlap zero are displayed in bold.

| PA artifact formula | Marg.Like | coeff1 | coeff2 |
|---|---|---|---|
| ~1 | -432.7 | - | - |
| ~1 + poly(rain,2) | -435.2 | **0.29–- 7.22** | -5.34–- 2.96 |
| ~1 + poly(RH,2) | -435.3 | **0.69–- 8.26** | -5.46–- 2.08 |
| ~1 + poly(Temp,2) | -437.5 | -4.95–- 1.55 | -0.97–- 4.9 |
| ~1 + RH | -437.8 | **0–- 0.06** | - |
| ~1 + poly(Forest,2) | -437.9 | -2.96–- 4.17 | -5.3–- 1.64 |
| ~1 + poly(TWI,2) | -438.0 | -5.25–- 1.88 | -3.43–- 3.91 |
| ~1 + rain | -438.2 | **0–- 0.03** | - |
| ~1 + Temp | -439.3 | -0.09–- 0.04 | - |
| ~1 + TWI | -440.2 | -0.03–- 0.01 | - |
| ~1 + Forest | -442.0 | **0–- 0.01** | - |

**Table 2.** Models exploring the effect of the environmental covariates on the occurrence intensity of *O.turicata americanus*. Poly refers to the quadratic function for the respective variable. The log of the marginal likelihood of each model is reported (Marg.Like), as well as the estimated credible intervals for each coefficient for the covariable (coeff1, coeff2). Coefficients that do not overlap zero are displayed in bold. The effect of the variable describing spatial sampling bias for each model is also reported.

|  | | Clim PC1 | | Clim PC2 | | Soil Composition | | Soil PC1 | | |
| --- | --- | --- | --- | --- | --- | --- | --- | --- | --- | --- |
| Distribution formula | Ma.Like | coeff1 | coeff2 | coeff1 | coeff2 | coeff1 | coeff2 | coeff1 | coeff2 | Samp. bias |
| ~ poly(Clim_PC1,2) + poly(Clim_PC2,2) + Soil_Comp | -430.2 | **0.87–-2.62** | **-1.9 - -0.86** | **-0.91 - -0.3** | **-1.48 - -0.5** | **-1.15 - -0.24** | - | - | - | -0.11–- 0.65 |
| ~ poly(Clim_PC1,2) + poly(Clim_PC2,2) + poly(Soil_Comp, 2) | -431.5 | **-2.2 - -0.21** | -0.31–- 0.87 | **-1.86 - -1.24** | **-2.86 - -1.86** | **-16.08 - -9.81** | **-6.52 - -3.86** | - | - | -1.31 - -0.5 |
| ~ 0 + poly(Clim_PC1,2) | -432.7 | **1.98–- 3.5** | **-2.34 - -1.34** | - | - | - | - | - | - | 0.03–- 0.76 |
| ~ 0 + poly( Soil_Comp,2) | -442.2 | - | - | - | - | **-9.78 - -4.1** | **-4.05 - -1.57** | - | - | 0.12–- 0.85 |
| ~ poly(Clim_PC2,2) | -455.2 | - | - | **-1.19 - -0.49** | **-1.78 - -0.75** | - | - | - | - | 0.56–- 1.25 |
| ~ Soil_Comp | -459.4 | - | - | - | - | **-3.29 - -1.02** | - | - | - | 0.22–- 0.9 |
| ~ 0 + Clim_PC1 | -464.5 | **0.45–- 1.89** | - | - | - | - | - | - | - | 0.52–- 1.2 |
| ~ Clim_PC2 | -469.4 | - | - | -1.34–- 0.4 | - | - | - | - | - | 0.67–- 1.37 |
| ~ Soil_PC1 | -469.7 | - | - | - | - | - | - | -0.58–- 0.96 | - | 0.63–- 1.36 |
| ~ poly(Soil_PC1,2) | -470.9 | - | - | - | - | - | - | **0.37–- 1.22** | -0.72–- 0.01 | 0.64–- 1.32 |

**Figures**

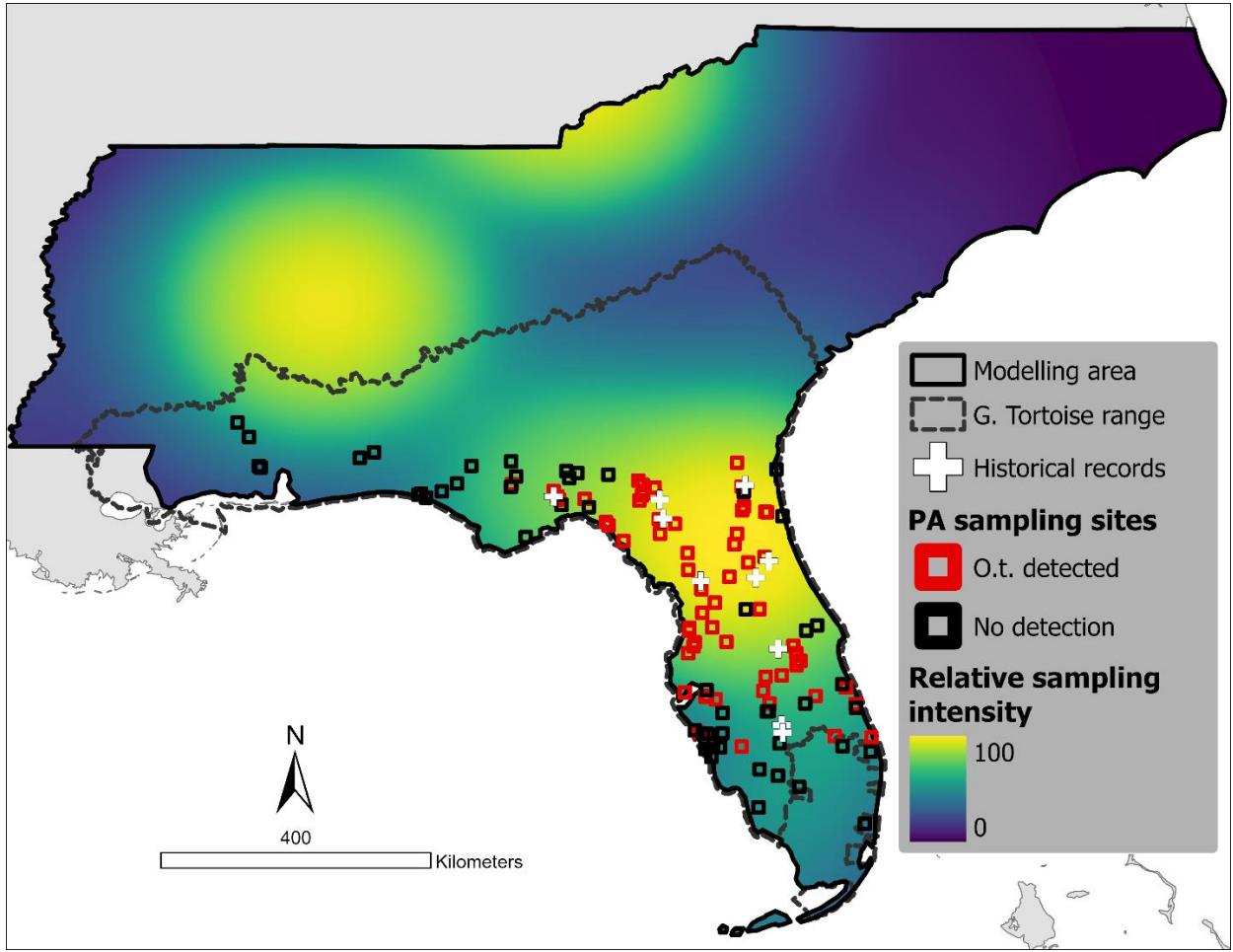

**Figure 1.** Modeling area showing the location of the systematic sampling sites (squares), with the sites where at least one gopher tortoise burrow provided a positive detection of *O. t. americanus* highlighted in red. Historical tick presence records georeferenced with a precision <15 km are also included as white crosses. The background displays the sampling bias, as estimated from the density of tick (Ixodida) occurrence records over the modeling area, as well as the range of the habitat host of *O.t. americanus*, the gopher tortoise *Gopherus polyphemus*.

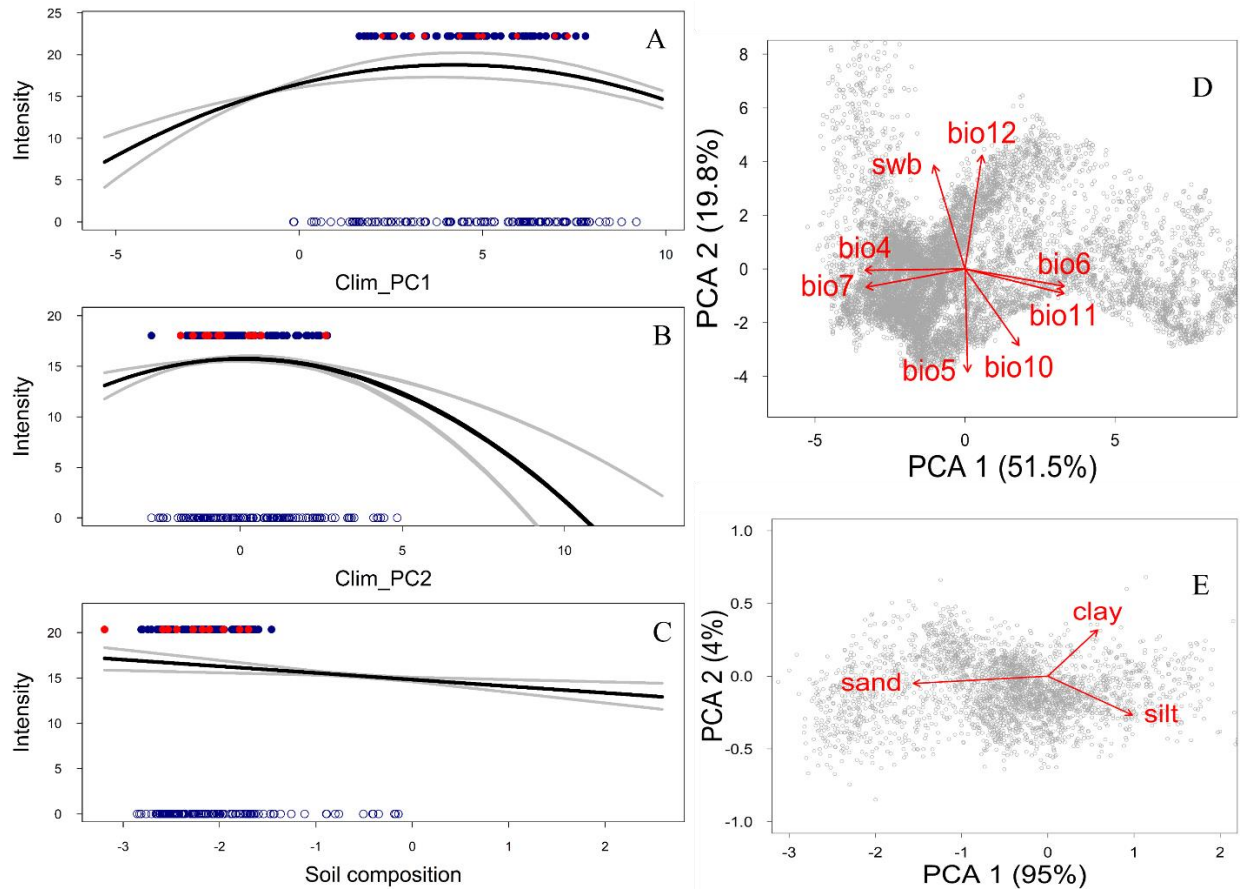

**Figure 2.** Environmental covariates and their effect on the occurrence of *O. t. americanus* in SE US. A) Effect of the first principal component of climatic conditions on the intensity of occurrence. B) Effect of the second principal component of climatic conditions. C) Effect of the robust component describing soil composition. D) Biplot of the first two principal components of climatic conditions showing the loadings of their most loaded variables: bio 4: Temperature Seasonality; bio 5: Max Temperature of Warmest Month; bio 6: Min Temperature of Coldest Month; bio 7: Temperature Annual Range; bio 10: Mean Temperature of Warmest Quarter; bio 11: Mean Temperature of Coldest Quarter; bio 12: Annual precipitation; swb: site water balance. E) Biplot of a robust PCA summarizing soil composition. Open points in A-C show sampled burrows with no soft ticks detected; blue filled points represent positive sampled burrows; and red dots show the historical records.

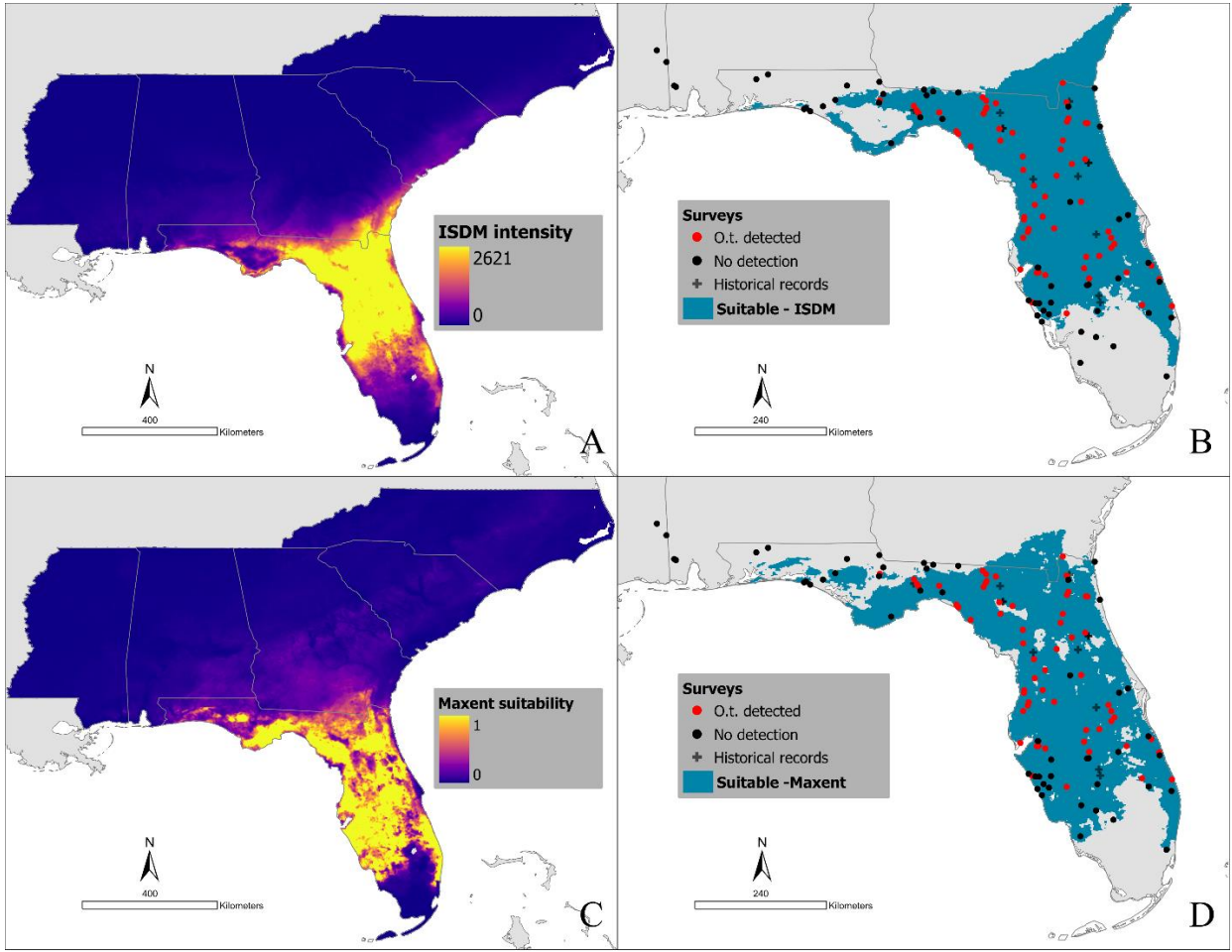

**Figure 3.** Spatial predictions of *Ornithodoros turicata americanus* distribution models. A) Predicted relative intensity of occurrence by the ISDM. B) Reclassification of the output of the ISDM to display predicted distribution using a threshold that keeps 98% of the systematically collected records. C) Suitability predicted by the Maxent model. D) Reclassification of the output of the Maxent model to display predicted distribution using a threshold that keeps 98% of all occurrence records.

36

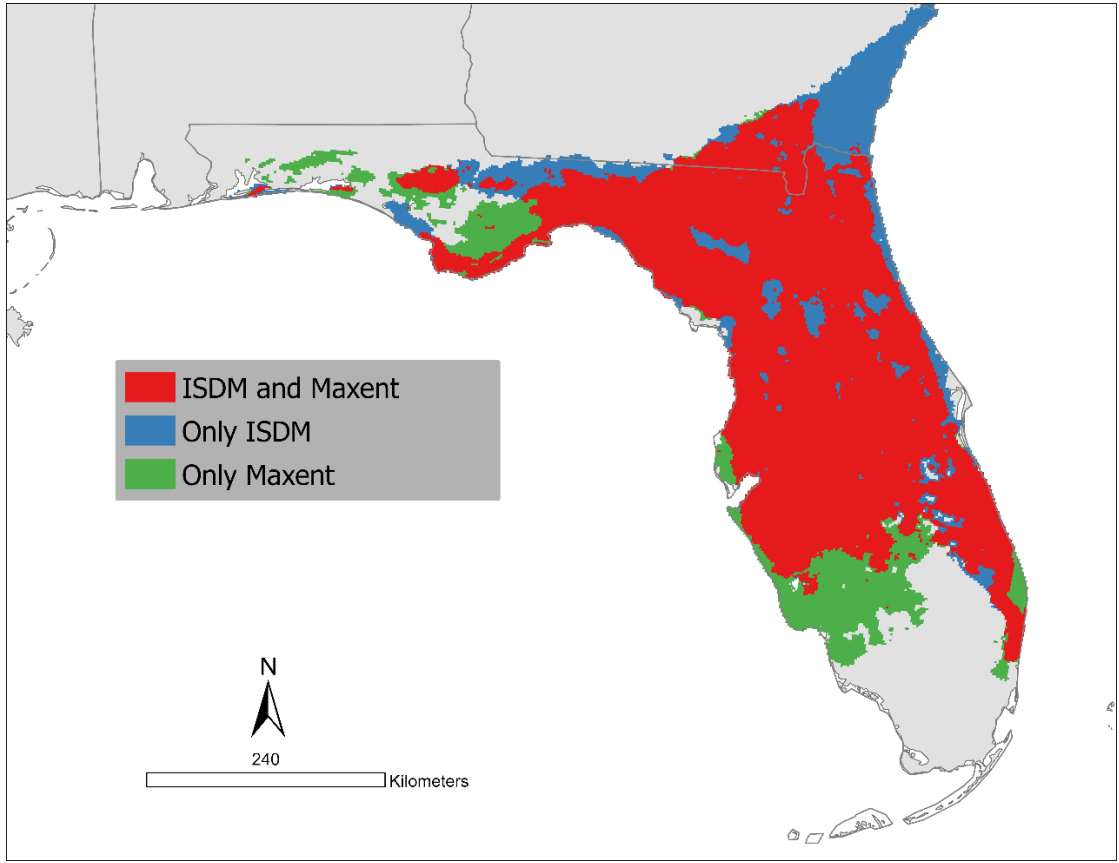

37

38 **Figure 4.** Concordance of the predicted *Ornithodoros turicata americanus* distribution by the
39 ISDM and Maxent models.

40

41

42

43

44

45

46

47

48

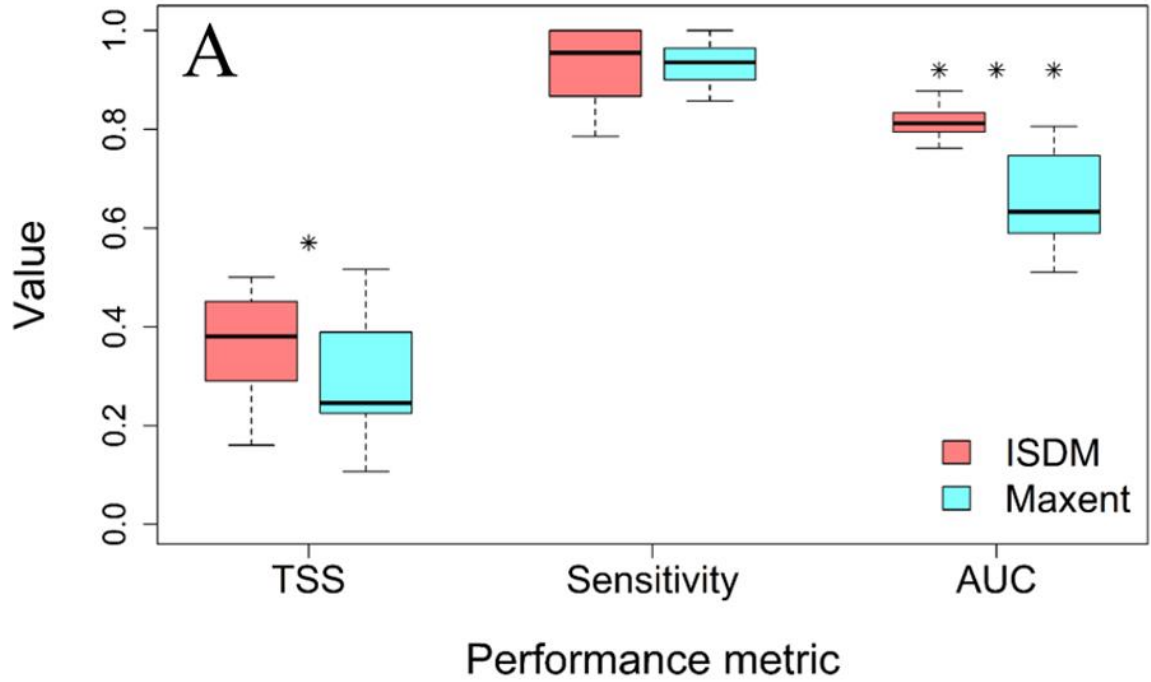

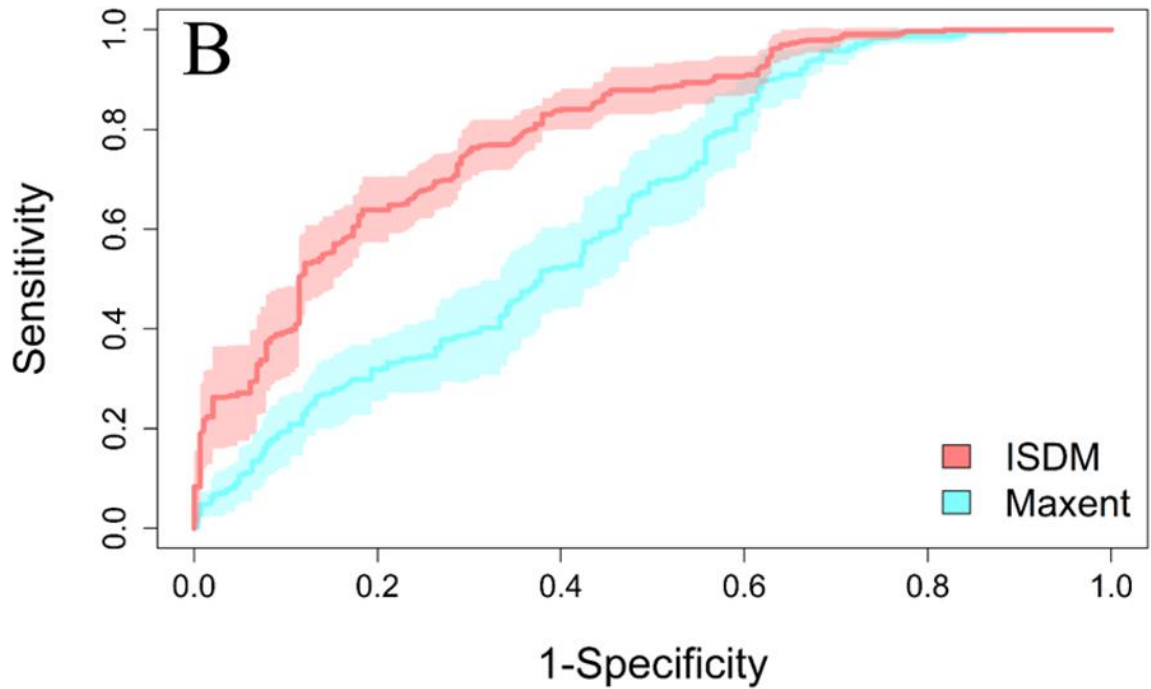

**Figure 5.** *Ornithodoros turicata americanus* distribution model evaluation and comparison. A) Boxplot comparing the model evaluation metrics estimated with a validation PA subset in ten model replicates for the ISDM and Maxent models. *Denotes significant differences (P-

53 value=0.05); *** Denotes highly significant differences (p-value=0.0002). B) Mean and
54 confidence intervals of the receiver-operator curves of the ten replicates for each model.